\documentclass{elsart}
\usepackage{ifpdf}
\usepackage{graphicx,amssymb}
\usepackage{amsbsy}
\usepackage{mathrsfs}

\newcommand{\eps}{\varepsilon}
\newcommand{\abs}[1]{\left\vert#1\right\vert}

\newcommand{\Pvs}{{\hbox{EEP}}}

\newcommand{\A}{A}
\newcommand{\B}{B}
\newcommand{\As}{a}
\newcommand{\Bs}{b}

\newcommand{\Zeta}{Z}

\begin{document}
\begin{frontmatter}
\title{On rotational solutions for elliptically excited pendulum}

\author{Anton O. Belyakov}
\address{Institute of Mechanics,
Moscow State Lomonosov University \\
Michurinsky pr. 1, Moscow 119192, Russia;\\
ORCOS, Institute of Mathematical Methods in Economics,\\ Vienna
University of Technology\\Argentinierstrasse 8/105-4, A-1040
Vienna, Austria} \ead{a\_belyakov@inbox.ru\\}

\begin{abstract}
The author considers the planar rotational motion of the
mathematical pendulum with its pivot oscillating both vertically
and horizontally, so the trajectory of the pivot is an ellipse
close to a circle. The analysis is based on the exact rotational
solutions in the case of circular pivot trajectory and zero
gravity. The conditions for existence and stability of such
solutions are derived. Assuming that the amplitudes of excitations
are not small while the pivot trajectory has small ellipticity the
approximate solutions are found both for high and small linear
damping. Comparison between approximate and numerical solutions is
made for different values of the damping parameter.
\end{abstract}
\end{frontmatter}

\section{Introduction}
\begin{figure}
  \center
  \includegraphics[width=0.4\textwidth]{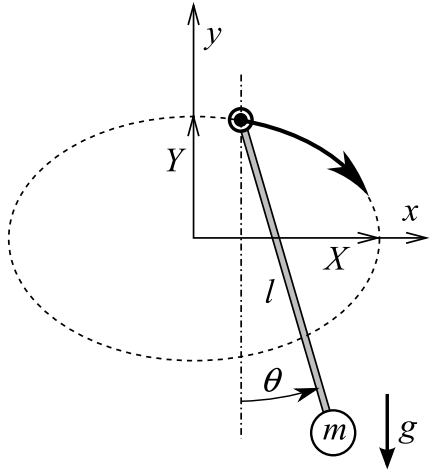}\\
  \caption{Scheme of the elliptically excited mathematical pendulum of length $l$. The pivot of the pendulum
  moves along the elliptic trajectory (dashed line) with semiaxis $X$ and $Y$.}\label{f:ElliptPendul}
\end{figure}

Elliptically excited pendulum ({\Pvs}) is a mathematical pendulum
in the vertical plane whose pivot oscillates not only vertically
but also horizontally with $\pi/2$ phase shift, so that the pivot
has elliptical trajectory, see Fig.~\ref{f:ElliptPendul}. {\Pvs}
is a natural generalization of pendulum with vertically vibrating
pivot that is one of the most studied classical systems with
parametric excitation, so it is often referred to simply as
parametric pendulum, see for example
\cite{Lenci_Rega,Xu_Wiercigroch,Bog_Mitr,Seyran} and references
therein.

Dynamics of {\Pvs} has been studied numerically and analytically
in \cite{Horton}. Approximate oscillatory and rotational solutions
for {\Pvs} are the common examples in literature
\cite{Blekhman54,Blekhman79,Blekhman,Akulenko} on asymptotic
methods. Sometimes {\Pvs} is presented in a slightly more general
model of unbalanced rotor \cite{Blekhman54,Blekhman79,Blekhman},
where the phase shift between vertical and horizontal oscillations
of the pivot can differ from $\pi/2$. {\Pvs} is also a special
case of generally excited pendulum in \cite{Trueba}.

The usual assumption for approximate solution in the literature is
the smallness of dimensionless damping and pivot oscillation
amplitudes in the {\Pvs}'s equation of motion. The author could
find only one paper \cite{Fidlin}, where oscillations of {\Pvs}
with high damping and yet small relative excitation were studied.

In the present paper we study rotations of {\Pvs} with not small
excitation amplitudes and with both small and not small linear
damping. Our analysis uses the exact solutions for {\Pvs} with the
absence of gravity and with equal excitation amplitudes, when
elliptical trajectory of the pivot becomes
circular.~\footnote{When there is no gravity the model of {\Pvs}
coincides with that of hula-hoop, see \cite{Belyak} and references
therein.}

The paper is organized as follows. In Section \ref{s:main} the
dimensionless equation of {\Pvs} motion is derived. In Section
\ref{s:exact} the exact rotational solutions and their stability
conditions are obtained in the case, with no gravity and the
circular trajectory of the pivot. In Section \ref{s:notsmall}
first and second order approximate solutions are obtained by
multiple scale method \cite{Nayfeh} for the close to circle
trajectory of the pivot and high damping, where we assume that
gravity is small or the frequency of excitation is high. In
Section \ref{s:small} for the same excitation and small damping
second order approximate solutions are obtained with the use of
averaging method \cite{Bog_Mitr,Vol_Morg}. In Section
\ref{s:domain} both solutions in Sections \ref{s:notsmall} and
\ref{s:small} are compared with the numerical solutions for
different values of the damping parameter.

\section{Main relations}\label{s:main}
Equation of {\Pvs}'s motion can be derived with the use of angular
momentum alteration theorem (e.~g. \cite{Horton})
\begin{equation}\label{eq:swing}
    m\,l^2 \frac{d^2 \theta}{d t^2} + c \frac{d \theta}{d t}
    + m\,l\left(g - \frac{d^2 y(t)}{d t^2}\right)\sin(\theta) - m\,l\,\frac{d^2 x(t)}{d t^2}\cos(\theta) = 0,
\end{equation}
where $l$ is the distance between the pivot and the concentrated
mass $m$; $c$ is the viscous damping coefficient; $\theta$ is the
angle of the pendulum deviation from the vertical position; $t$ is
time; $g$ is gravitational acceleration.

It is assumed that the pivot of the pendulum moves according to
the periodic law
\begin{equation}\label{eq:y}
    x = X \sin(\Omega t), \quad y = Y \cos(\Omega t),
\end{equation}
where $X$, $Y$, and $\Omega$ are the amplitudes and frequency of
the excitation.

We introduce the following dimensionless parameters and new time
\begin{eqnarray}\label{eq:param}
&&\Omega_0 = \sqrt{\frac{g}{l}},\quad \eps =
\frac{Y-X}{2\,l},\quad \mu = \frac{Y+X}{2\,l}>0,\nonumber\\
&&\omega = \frac{\Omega_0}{\Omega},\quad \beta = \frac{c}{
m\,l^2\,\Omega},\quad \tau=\Omega t,
\end{eqnarray}
in which equation (\ref{eq:swing}) with substituted (\ref{eq:y})
in it takes the following form
\begin{equation}\label{eq:theta}
\ddot{\theta} + \beta \dot{\theta} + \mu\sin\!\left( \tau+\theta
\right) = \eps\sin\!\left( \tau-\theta \right) - \omega^2
\sin(\theta),
\end{equation}
were we use formula $Y\cos\!\left(\Omega\,t\right)\sin\!\left(
\theta \right) +X\sin\!\left( \Omega\,t \right) \cos\!\left(
\theta \right) = \frac{Y+X}{2}\sin\!\left( \Omega\,t+\theta
\right) - \frac{Y-X}{2}\sin\!\left( \Omega\,t-\theta \right)$.
Here the upper dot denotes differentiation with respect to new
time $\tau$.

\section{Exact rotational solution when $\eps=0$ and $\omega =
0$}\label{s:exact} Conditions $\eps=\omega=0$ mean that we find
the mode of rotation for the circular excitation $X = Y$ with
absence of gravity $g = 0$. In this case, we call equation
(\ref{eq:theta}) the \emph{unperturbed equation}
\begin{equation}\label{eq:theta0}
\ddot{\theta} + \beta\dot{\theta} + \mu\sin\!\left(
\tau+\theta\right) = 0
\end{equation}
which has exact solutions
\begin{equation}\label{eq:sol0}
\theta = \theta_0 - \tau,
\end{equation}
where constants $\theta_0$ are defined by the following equality
\begin{equation}\label{eq:sinpsi}
\sin(\theta_0) = \frac{\beta}{\mu},
\end{equation}
provided that $\abs{\beta}\leq \mu$.

To investigate the stability of these solutions we present the
angle $\theta$ as $\theta = \theta_0 - \tau + \eta$, where
$\eta=\eta(\tau)$ is a small addition, and substitute it in
equation (\ref{eq:theta0}). Then linearizing in (\ref{eq:theta0})
and using equality (\ref{eq:sinpsi}), we obtain the linear
equation
\begin{equation}\label{eq:eta}
\ddot\eta + \beta\dot\eta + \mu\cos(\theta_0)\eta = 0,
\end{equation}
According to the Lyapunov stability theorem, solution
(\ref{eq:sol0}) is asymptotically stable according to the linear
approximation if all eigenvalues of linearized equation
(\ref{eq:eta}) have negative real parts. Which happens when the
following inequalities are satisfied
\begin{equation}\label{eq:stab}
\beta > 0, \quad \mu\cos(\theta_0) > 0,
\end{equation}
obtained from the Routh--Hurwitz conditions. From conditions
(\ref{eq:stab}), assumption $\mu
> 0$ in (\ref{eq:param}), and equality (\ref{eq:sinpsi}), it follows for
$\beta > 0$ that the solutions
\begin{equation}\label{eq:sol01}
\theta = \theta_0 - \tau, \quad \theta_0 =
\arcsin\!\left(\frac{\beta}{\mu}\right) + 2 \pi k
\end{equation}
are asymptotically stable, while the solutions
\begin{equation}\label{eq:sol02}
\theta = \theta_0 - \tau, \quad \theta_0 =
\pi-\arcsin\!\left(\frac{\beta}{\mu}\right) + 2 \pi k
\end{equation}
are unstable, where $k$ is any integer number. For negative
damping, $\beta < 0$, both these solutions are unstable. From now
on we will assume that the following conditions are satisfied
\begin{equation}\label{eq:existab}
0 < \beta < \mu,
\end{equation}
which ensure the existence of stable rotational solution
(\ref{eq:sol01}) as it is seen from (\ref{eq:sinpsi}) and
(\ref{eq:stab}). Indeed, in order to guarantee asymptotic
stability $\beta$ should be not only positive, but also strictly
less than $\mu$ because of the second condition in
(\ref{eq:stab}), which can be transformed to inequality
$\mu\cos(\theta_0) = \sqrt{\mu^2-\beta^2} > 0$ with the use of the
positive root for $\mu\cos(\theta_0)$ from (\ref{eq:sinpsi}).

\section{Approximate rotational solutions when $\eps\approx 0$
and $\omega \sim \sqrt{\eps}$}\label{s:notsmall} We assume that
values of $\eps$ and $\omega^2$ are small of the same order of
smallness, i.e. $\eps \sim \omega^2 \ll 1$, so we can introduce
new parameter $w = \omega^2/\eps$.

One can deduct from (\ref{eq:param}) and current assumptions that
either gravity $g$ is small or the frequency of excitation
$\Omega$ is high with such damping $c$ and mass $m$ so that
damping coefficient $\beta\sim 1$.

All small terms are in the right-hand side of equation
(\ref{eq:theta}). To solve equation (\ref{eq:theta}) we will use
multiple scale method \cite{Nayfeh}. In this method general
solution of equation (\ref{eq:theta}) is assumed to be of the
following form
\begin{equation}\label{eq:gensol}
\theta = - \tau + \theta_0 + \eps\theta_1 + \eps^2\theta_2 +
\ldots
\end{equation}
A series of time scales (independent variables), $T_0$, $T_1$,
$\ldots$, is introduced, where $T_0 = \tau$, $T_1 = \eps\tau$,
$\ldots$. So that $\theta$ is a function of these time scales,
$\theta(T_0, T_1, \ldots)$. Using the chain rule, the time
derivatives become
\begin{eqnarray}
  \frac{d}{d\tau} &=& D_0 + \eps D_1 + \eps^2 D_2 + \ldots \label{eq:td1}\\
  \frac{d^2}{d\tau^2} &=& D_0^2 + 2\eps D_0 D_1 + \eps^2 (D_0 D_2 + D_1^2) + \ldots \label{eq:td2}
\end{eqnarray}
where $D_n^m = \partial^m/\partial T_n^m$. Next the general
solution and the time derivatives are substituted into equation
(\ref{eq:theta}), where sines are expended into the Taylor series
with respect to $\eps$. By grouping together the terms with the
same powers of $\eps$ and equating to zero, a set of differential
equations is obtained,
\begin{eqnarray}
 & & D_0^2\theta_0 + \beta D_0 (\theta_0 - T_0) + \mu\sin(\theta_0)  =  0, \label{eq:D0}\\
 & & D_0^2\theta_1 + \beta D_0 \theta_1  + \mu\cos(\theta_0)\theta_1 = - (2 D_0
 D_1 + \beta D_1)(\theta_0-T_0)\nonumber\\
&+&\sin\!\left( 2T_0-\theta_0\right) + w\sin(T_0
-\theta_0), \label{eq:D1}\\
& & D_0^2\theta_2 + \beta D_0 \theta_2 + \mu\cos(\theta_0)\theta_2
= \mu\sin(\theta_0)\theta_1^2 / 2 - (2 D_0
 D_1 + \beta D_1)\theta_1 \nonumber\\
 & - & (D_0 D_2 + D_1^2 + \beta D_2)(\theta_0 - T_0)\nonumber\\
 & - & \left(\cos\!\left( 2T_0-\theta_0\right) + w\cos(T_0
-\theta_0)\right)\theta_1,\label{eq:D2}\\
& \ldots &\nonumber
\end{eqnarray}
where we denote  $w = \omega^2/\eps$. We have already found
solution (\ref{eq:sol0}) for equation (\ref{eq:D0}) in the
previous section. Here we consider the same stable regular
rotations 1:1 whose zero approximation is given by
(\ref{eq:sol01}). Hence, $\theta_0$ is a constant and consequently
we have $(2 D_0
 D_1 + \beta D_1)(\theta_0-T_0) = 0$ and $(D_0 D_2 + D_1^2 + \beta D_2)(\theta_0 - T_0) =
 0$. Thus, equations (\ref{eq:D1}) and (\ref{eq:D2}) can be
 written in the following way
\begin{eqnarray}
 & & D_0^2\theta_1 + \beta D_0 \theta_1  + \sqrt{\mu^2-\beta^2}\theta_1 = \sin\!\left( 2T_0-\theta_0\right) + w\sin(T_0
-\theta_0) \label{eq:d1}\\
& & D_0^2\theta_2 + \beta D_0 \theta_2 +
\sqrt{\mu^2-\beta^2}\theta_2 = \beta \theta_1^2 / 2 - \left(2 D_0
 D_1 + \beta D_1\right)\theta_1 \nonumber\\
 & - & \left(\cos\!\left( 2T_0-\theta_0\right) + w\cos(T_0
-\theta_0)\right)\theta_1,\label{eq:d2}
\end{eqnarray}
where we denote $\mu\sin(\theta_0) = \beta$ and $\mu\cos(\theta_0)
= \sqrt{\mu^2-\beta^2}$ with the use of relation (\ref{eq:sinpsi})
and the second condition in (\ref{eq:stab}).

\subsection{First order approximation}
In consequence of conditions (\ref{eq:existab}) non-homogeneous
linear differential equation (\ref{eq:d1}) can be presented in the
following form
\begin{eqnarray}
 & &  D_0^2\theta_1 + \beta D_0 \theta_1  + \sqrt{\mu^2-\beta^2}\theta_1 \nonumber\\
 & = & \A_1\cos\!\left( T_0\right) + \B_1\sin\!\left( T_0\right)\nonumber\\
 & + & \A_2\cos\!\left( 2T_0\right) + \B_2\sin\!\left( 2T_0\right), \label{eq:sd1}
\end{eqnarray}
where $\A_1 = -w\beta/\mu$, $\B_1 = w\sqrt{1-\beta^2/\mu^2}$,
$\A_2 = -\beta/\mu$, $\B_2 = \sqrt{1-\beta^2/\mu^2}$, lower index
denotes harmonic number. Equation (\ref{eq:sd1}) has a unique
periodic solution
\begin{eqnarray}
 \theta_1(T_0) & = & \As_1\cos\!\left( T_0\right) + \Bs_1\sin\!\left( T_0\right)\nonumber\\
 & + & \As_2\cos\!\left( 2T_0\right) + \Bs_2\sin\!\left( 2T_0\right), \label{eq:theta1}
\end{eqnarray}
where $\As_1 = -\frac {(1-\sqrt {{\mu}^{2}-\beta^{2}})\A_1 +
  \beta\B_1}{{\mu}^{2}+1-2\sqrt{\mu^2-\beta^2}}$, $\Bs_1 = -
\frac{-\beta\A_1 +
(1-\sqrt{\mu^2-\beta^2})\B_1}{{\mu}^{2}+1-2\sqrt{\mu^2-\beta^2}}$,
$\As_2 = -\frac {(4-\sqrt{{\mu}^{2}-\beta^{2}})\A_2 + 2\beta\B_2
}{3\beta^{2}+{\mu}^{2}+4(4-2\sqrt{\mu^2-\beta^2})}$, and $\Bs_2 =
\frac{-2\beta\A_2 +
(4-\sqrt{\mu^2-\beta^2})\B_2}{3\beta^{2}+{\mu}^{2}+4(4-2\sqrt{\mu^2-\beta^2})}$.

Thus, the solution for (\ref{eq:theta}) in the first approximation
can be written as follows
\begin{eqnarray}
 \theta & = & -\tau + \theta_0\nonumber\\
 & & - \eps\,\frac{2\,\beta\cos\!\left( 2 T_0-\theta_0 \right) +
 \left(4 - \sqrt{\mu^2-\beta^2}\right) \sin\!\left( 2 T_0 -
 \theta_0 \right)}{3\beta^{2}+\mu^{2}+8(2-\sqrt{\mu^2-\beta^2})}\nonumber\\
 & & - \omega^2\,\frac{\beta\cos\!\left(T_0-\theta_0 \right) +
 \left(1-\sqrt{\mu^2-\beta^2}\right)
 \sin\!\left(T_0-\theta_0\right)}{\mu^2+1-2\,\sqrt{\mu^2-\beta^2}},\label{eq:stheta1}
\end{eqnarray}
where constant $\theta_0$ is defined in (\ref{eq:sol01}).

\subsection{Second order approximation}
\begin{figure}
  \includegraphics[width=0.9\textwidth]{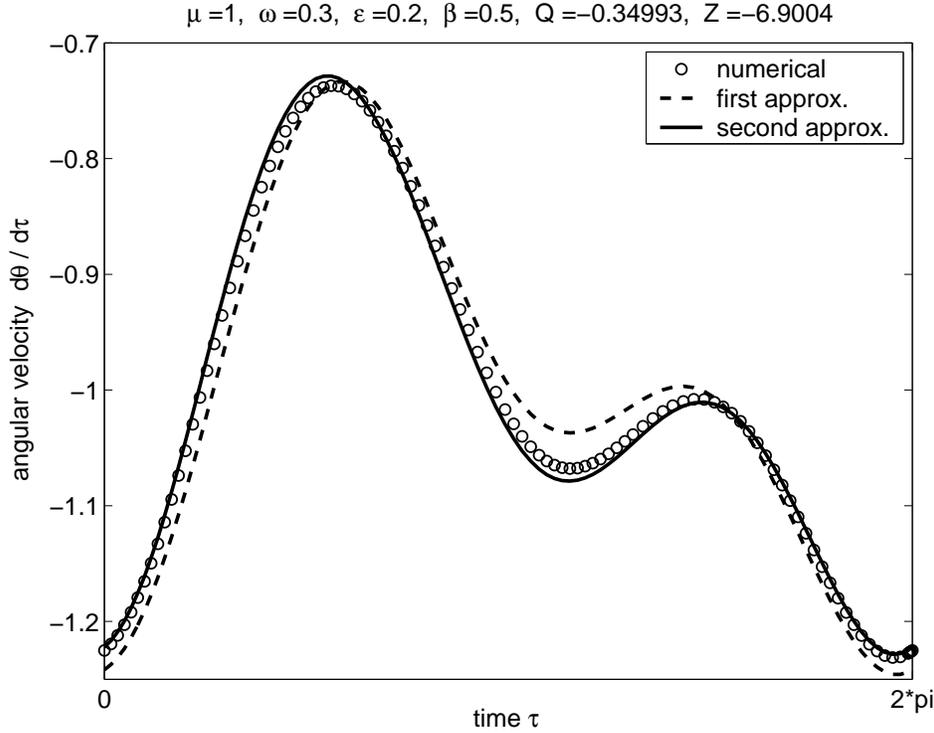}\\
  \caption{Angular velocities $\dot{\theta}$ calculated from
  the first order approximate solution (\ref{eq:stheta1}), second order approximate solution (\ref{eq:stheta2}), and results of numerical
  simulation, when damping coefficient $\beta$ is not small.}\label{f:B05}
\end{figure}

Since (\ref{eq:theta1}) does not contain any constant of
integration we set $(2 D_0
 D_1 + \beta D_1)\theta_1 = 0$ in equation (\ref{eq:d2}) and substitute in it an expression $\cos\!\left( 2T_0-\theta_0\right) +
w\cos(T_0 -\theta_0) = \B_1\cos\!\left(T_0\right) -\A_1\sin\!
\left(T_0\right) +\B_2\cos\!\left( 2 T_0 \right) -\A_2\sin\!\left(
2 T_0 \right)$ with coefficients defined in (\ref{eq:sd1}). Thus,
equation (\ref{eq:sd1}) takes the following form
\begin{equation}\label{eq:dd2}
D_0^2\theta_2 + \beta D_0 \theta_2 + \sqrt{\mu^2-\beta^2}\theta_2
= \frac{\A_0'}{2}+\sum_{n=1}^4 \left(\A_n'\cos\!\left(n T_0\right)
+ \B_n'\sin\!\left(n T_0\right)\right),
\end{equation}
where coefficients in the right-hand side are the following
\begin{eqnarray}
 \A_0' &=& \left({\Bs_{{2}}}^{2}+{\Bs_{{1}}}^{2}+{\As_{{2}}}^{2}+{\As_{{1}}}^{2} \right)\beta+(\A_{{1}}\Bs_{{1}}+\A_{{2}}\Bs_{{2}}-\B_{{2}
}\As_{{2}}-\B_{{1}}\As_{{1}})\\
 \A_1' &=& \left( \As_{{1}}\As_{{2}}+\Bs_{{1}}\Bs_{{ 2}} \right) \beta+(\A_1\Bs_2+\A_2\Bs_1-\B_1\As_2-\B_2\As_1)/2\\
 \A_2' &=& \left( \As_1^{2}-\Bs_1^{2}
 \right)\beta/2-(\A_1\Bs_1+\B_1\As_1)/2\\
 \A_3' &=& \left( \As_{{1}}\As_{ {2}} -\Bs_{{1}}\Bs_{{2}}\right)
\beta-(\A_{{2}}\Bs_{{1}}+\A_1\Bs_2+\B_{{1}}\As_{{2
}}+\B_{{2}}\As_{{1}})/2\\
 \A_4' &=& \left(\As_2^{2}-\Bs_2^{2}
\right) \beta/2-(\A_2\Bs_2+\B_2\As_2)/2\\
\B_1' &=& \left(\As_{{1}}\Bs_{{2}}-\Bs_{{1}}\As_{{2}}\right)
\beta-(\A_1\As_2-\A_2\As_1+\B_1\Bs_2-\B_2\Bs_1)/2\\
 \B_2' &=& \beta\,\As_{{1}}\Bs_{{1}}+(\A_1\As_1-\B_1\Bs_1)/2\\
 \B_3' &=& \left( \As_{{1}}\Bs_{{2}}+\Bs_{{1}}\As_{{ 2}} \right)
\beta+(\A_1\As_2+\A_2\As_1-\B_1\Bs_2-\B_2\Bs_1)/2\\
 \B_4' &=& \beta\,\As_{{2}}\Bs_{{2}}
+(\A_2\As_2-\B_2\Bs_2)/2.
\end{eqnarray}
Periodic solution for equation (\ref{eq:dd2}) has the following
form which is obtained from (\ref{eq:gensol}) in Appendix
\begin{eqnarray}
  \theta_2(T_0) & = &  \frac{\A_0'}{2\sqrt{{\mu}^{2}-\beta^{2}}}\nonumber\\
  &-&\sum_{n=1}^4 \frac {(n^2-\sqrt {{\mu}^{2}-\beta^{2}})\A_n' +
  n\beta\B_n'
}{(n^2-1)\beta^{2}+{\mu}^{2}+n^2(n^2-2\sqrt{\mu^2-\beta^2})}\cos\!\left(n T_0\right)\nonumber\\
 & - & \sum_{n=1}^4\frac{-n\beta\A_n' + (n^2-\sqrt{\mu^2-\beta^2})\B_n'}{(n^2-1)\beta^{2}+{\mu}^{2}+n^2(n^2-2\sqrt{\mu^2-\beta^2})}\sin\!\left(
n T_0 \right),\label{eq:theta2}
\end{eqnarray}
where constant term is derived from (\ref{eq:gensol0}) taking
$\A_0 = \A_0'/2$. Thus, second order approximate solution can be
shortly written in the following form
\begin{eqnarray}
  \theta & = & -\tau + \theta_0 + \eps\theta_1(\tau)+\eps^2\theta_2(\tau), \label{eq:stheta2}
\end{eqnarray}
where constant $\theta_0$ is defined in (\ref{eq:sol01}), function
$\theta_1$ in (\ref{eq:theta1}), and function $\theta_2$ in
(\ref{eq:theta2}). In Fig. \ref{f:B05} it is shown how first and
second order approximate solutions approach the numerical
solution.

In this section we have used the fact that all solutions in each
time scale must converge to corresponding unique periodic
solutions because damping $\beta$ is not small. If it is not the
case the multiple scale analysis becomes more complicated. In the
next section we will tackle the problem of small damping
$\beta\sim\sqrt{\eps}$ with the use of classical averaging
technique.

\section{Approximate rotational solutions when $\eps\approx 0$, $\omega \sim \sqrt{\eps}$, and $\beta\sim
\sqrt{\eps}$}\label{s:small} One can see in (\ref{eq:param}) that
assumptions $\omega \sim \beta \sim \sqrt{\eps}$ are valid for the
high frequency of excitation $\Omega\sim 1/\sqrt{\eps}$ with other
parameters being of order 1. Another option is small gravity
$g\sim \eps$ along with small ratio $c/m \sim \sqrt{\eps}$.

After change of variable $\theta = - \tau + \sqrt{\eps}\vartheta$
equation (\ref{eq:theta}) takes the following form
\begin{eqnarray}
& & \ddot{\vartheta} + \mu\vartheta - \hat{\beta} =
\mu\left(\vartheta - \frac{
 \sin\!\left(\sqrt{\eps}\vartheta\right)}{\sqrt{\eps}}\right)-\sqrt{\eps}\hat{\beta}\dot{\vartheta}\nonumber\\
 & + & \sqrt{\eps}\sin\!\left(2\tau-\sqrt{\eps}\vartheta\right) + \sqrt{\eps} w \sin(\tau - \sqrt{\eps}\vartheta),\label{eq:vartheta2}
\end{eqnarray}
with small right-hand side, where we denote $\hat{\beta} =
\beta/\sqrt{\eps}$ and as in the previous section $w =
\omega^2/\eps$. With zero right-hand side equation
(\ref{eq:vartheta2}) $\ddot{\vartheta} + \mu\vartheta -
\hat{\beta} = 0$ would describe harmonic oscillations about
$\hat{\beta}/\mu$ value with frequency $\sqrt{\mu}$. After
Taylor's expansion of sines in the right-hand side of
(\ref{eq:vartheta2}) about $\vartheta = 0$ we obtain the following
equation
\begin{eqnarray}
& & \ddot{\vartheta}  +  \left(\mu + \eps\cos\!\left(2\tau\right)
+ \eps w\cos\!\left(\tau\right)\right)\vartheta - \hat{\beta}  \nonumber\\
& = & \sqrt{\eps}\sin\!\left(2\tau\right)+\sqrt{\eps}
w\sin\!\left(\tau\right)-\sqrt{\eps}\hat{\beta}\dot{\vartheta}
  +\eps\mu\frac{\vartheta^3}{6} + o(\eps),\label{eq:vartheta3}
\end{eqnarray}
which describes oscillator with both basic and parametric
excitations. To solve equation (\ref{eq:vartheta3}) we will use
averaging method \cite{Bog_Mitr,Vol_Morg}. For that purpose we
will write (\ref{eq:vartheta3}) in the \emph{standard form} of
first order differential equations with small right-hand sides.
First, we use \emph{Poincar\'{e} variables} $q$ and $\psi$ defined
via the following solution of \emph{generating system}
$\ddot{\vartheta} + \mu\vartheta - \hat{\beta} = 0$ which is
(\ref{eq:vartheta3}) with $\eps=0$
\begin{equation}\label{eq_apsi}
\vartheta = \frac{\hat{\beta}}{\mu} + q \cos(\psi), \quad
\dot{\vartheta} = -\sqrt{\mu} q\sin(\psi).
\end{equation}
In Poincar\'{e} variables equation (\ref{eq:vartheta3}) becomes a
system of first order differential equations
\begin{eqnarray}
\dot q & = & - \frac{\sin\psi}{\sqrt{\mu}}f(\tau,q,\psi),\label{eq:dapsi}\\
\dot\psi & = & \sqrt{\mu} -
\frac{\cos\psi}{q\sqrt{\mu}}f(\tau,q,\psi),\label{eq:dpsi}
\end{eqnarray}
where small function $f(\tau,q,\psi) =
\sqrt{\eps}f_1(\tau,q,\psi)+\eps f_2(\tau,q,\psi)+o(\eps)$ is the
right hand side of (\ref{eq:vartheta2}), where
\begin{eqnarray}
f_1(\tau,q,\psi) & = & \sin\!\left(2\tau\right) + w
\sin\!\left(\tau\right) +\hat{\beta} q\sqrt{\mu}\sin(\psi),\label{eq:f1}\\
f_2(\tau,q,\psi) & =
& - \left(\cos\!\left(2\tau\right) + w
\cos\!\left(\tau\right)\right)\left(\frac{\hat{\beta}}{\mu} + q
\cos(\psi)\right)
\nonumber\\
& & +\frac{\mu}{6}\left(\frac{\hat{\beta}}{\mu} + q
\cos(\psi)\right)^3,\label{eq:f2}
\end{eqnarray}
meaning that $f(\tau,q,\psi)=O(\sqrt{\eps})$. Our next assumption
is that $\sqrt{\mu} - 1 \sim \sqrt{\eps}$ which means that
excitation frequency is close to the first resonant frequency of
basic excitation component $\sin\!\left(\tau\right)$ and to the
first resonant frequency of parametric excitation component
$\cos\!\left(2\tau\right)$ in equation (\ref{eq:vartheta3}). Thus,
system (\ref{eq:dapsi}), (\ref{eq:dpsi}) is transformed by $\psi =
\zeta + \tau$ to the standard form
\begin{eqnarray}
\dot{q} & = & -
\frac{1}{\sqrt{\mu}}\sin\!\left(\zeta+\tau\right)f\!\left(\tau,q,\zeta+\tau\right),\label{eq:dazeta}\\
\dot{\zeta} & = & \sqrt{\mu} - 1 -
\frac{1}{q\sqrt{\mu}}\cos\!\left(\zeta+\tau\right)f\!\left(\tau,q,\zeta+\tau\right),\label{eq:dzeta}
\end{eqnarray}
with small right-hand side, where new slow variable $\zeta$ is
often referred to as \emph{phase mismatch}.

\subsection{First order approximation}
In the first approximation so called \emph{averaged equations} can
be obtained by averaging the system (\ref{eq:dazeta}),
(\ref{eq:dzeta}) over period $2\pi$
\begin{eqnarray}
\dot{Q} & = & -
\frac{\sqrt{\eps}}{2\pi\sqrt{\mu}}\int^{2\pi}_0\sin\!\left(\Zeta+\tau\right)f_1\!\left(\tau,Q,\Zeta+\tau\right)\d\tau + o(\sqrt{\eps}),\label{eq:dQ}\\
\dot{\Zeta} & = & \sqrt{\mu} - 1 - \frac{\sqrt{\eps}}{2\pi
\sqrt{\mu}
Q}\int^{2\pi}_0\cos\!\left(\Zeta+\tau\right)f_1\!\left(\tau,Q,\Zeta+\tau\right)\d\tau
+ o(\sqrt{\eps}),\label{eq:dZeta}
\end{eqnarray}
where $Q$ and $\Zeta$ are the \emph{averaged variables}
corresponding to $q$ and $\zeta$. After taking the integrals we
have the following system
\begin{eqnarray}
\dot{Q} & = & -
\frac{\sqrt{\eps}w}{2\sqrt{\mu}}\cos\!\left(\Zeta\right) - \frac{\sqrt{\eps}\hat{\beta}}{2}Q + o(\sqrt{\eps}),\label{eq:dQ1}\\
\dot{\Zeta} & = & \sqrt{\mu} - 1  +
\frac{\sqrt{\eps}w}{2\sqrt{\mu} Q}\sin\!\left(\Zeta\right) +
o(\sqrt{\eps}),\label{eq:dZeta1}
\end{eqnarray}
stationary solutions ($\dot{Q} = 0$, $\dot{\Zeta}=0$) of which are
the following
\begin{eqnarray}
  Q^2 & = & \frac{\omega^2/\eps}{\mu\,(4(\sqrt{\mu} - 1)^2 + \beta^2)}+ o(1),\\
  \Zeta & = & \arctan\!\left(\frac{2(\mu-1)}{\beta}\right) + 2\pi k + o(1),
\end{eqnarray}
where we have substituted back $w=\omega^2/\eps$ and
$\hat{\beta}=\beta/\sqrt{\eps}$. Symbol $\arctan$ stands for the
principal value of the function on the interval from $0$ to $\pi$.
Note that the phase $\Zeta$ is determined to within $2\pi$ rather
than $\pi$, since the functions $\sin(\Zeta)$ and $\cos(\Zeta)$
obtained from equations (\ref{eq:dQ1}) and (\ref{eq:dZeta1})
determine $\Zeta$ up to an additive term $2\pi k$. Solution of
system (\ref{eq:dazeta}-\ref{eq:dzeta}) in the first approximation
is $q = Q + o(1)$, $\zeta = \Zeta + o(1)$ so the solution of
(\ref{eq:theta}) is the following
$$\theta = -\tau +  \frac{\beta}{\mu} + \sqrt{\eps} Q \cos(\Zeta +
\tau) + o(\sqrt{\eps}),$$ which does not contain higher harmonics
observed numerically. That is why we need to proceed to the second
order approximation.

\subsection{Second order approximation}
\begin{figure}
  \includegraphics[width=0.9\textwidth]{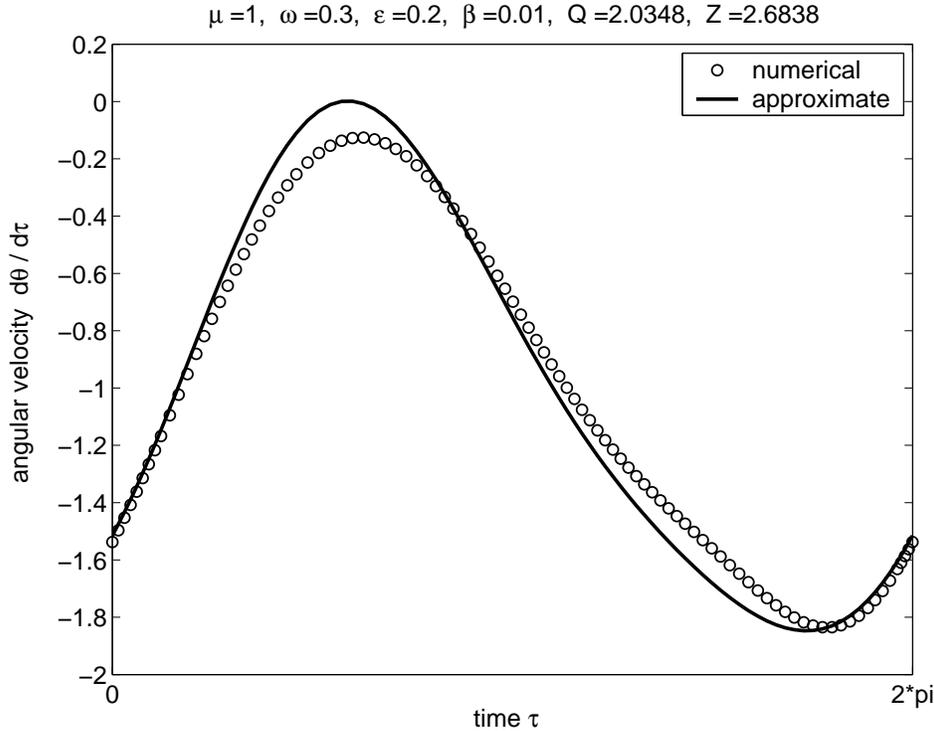}\\
  \caption{Angular velocity $\dot{\theta}$
  of the second order approximate solution (\ref{eq:thetaappr2}) compared with the results of numerical
  simulatios in the case of small damping $\beta$}\label{f:B001}
\end{figure}

In the second approximation averaged equations can be obtained as
follows
\begin{eqnarray}
\dot{Q} & = & \left(-\sqrt{\eps}\cos\!\left(\Zeta\right)
+\frac{\eps\hat{\beta}}{4}\left(\frac{4}{\mu} -
1\right)\sin\!\left(Z\right)\right)\frac{w}{2\sqrt{\mu}} \nonumber\\
  & & + \left(-\frac{\sqrt{\eps}\hat{\beta}}{2} + \frac{\eps}{4\sqrt{\mu}}\sin\!\left(2 Z \right)\right)Q + o(\eps),\label{eq:dQ2}\\
\dot{\Zeta} & = & \sqrt{\mu} - 1  +
\left(\sqrt{\eps}\sin\!\left(\Zeta\right) + \frac{\eps\hat{\beta}}{4}\left(\frac{4}{\mu} - 1\right)\cos\!\left(Z\right)\right)\frac{w}{2\sqrt{\mu} Q}\nonumber\\
 &  & - \frac{\eps\hat{\beta}^2}{8}\left(\frac{2}{\mu\sqrt{\mu}}+1\right) +
 \frac{\eps}{4\sqrt{\mu}}\cos\!\left( 2Z
 \right) - \frac{\eps\sqrt{\mu}}{16}Q^2 + o(\eps),\label{eq:dZeta2}
\end{eqnarray}
stationary solutions ($\dot{Q}=0$, $\dot{\Zeta}=0$) can be found
numerically or with absence of gravity ($\omega = 0$)
analytically. Solution of system (\ref{eq:dazeta}-\ref{eq:dzeta})
in the second approximation is the following
\begin{eqnarray}
  q & = & Q+\frac{\sqrt{\eps}}{2\sqrt{\mu}}\left(-\sin\!\left(\tau-Z\right)+\frac{w}{2}\sin\!\left(2\tau+Z\right)+\frac{1}{3}\sin\!\left(3\tau+Z\right)\right)\nonumber\\
& &+\sqrt{\eps}\frac{\hat{\beta}\,Q}{4}\sin\left(2\tau+2Z\right)+ o(\sqrt{\eps}),\\
  \zeta & = &\Zeta + \frac
{\sqrt{\eps}}{2\sqrt{\mu} Q}\left(\cos\!\left(\tau-Z\right)+\frac{w}{2}\cos\!\left(2\tau+Z\right)+\frac{1}{3}\cos\!\left(3\tau+Z\right)\right)\nonumber\\
&&
+\sqrt{\eps}\frac{\hat{\beta}}{4}\cos\left(2\tau+2Z\right)+o(\sqrt{\eps}),
\end{eqnarray}
Substitution of these expressions into (\ref{eq_apsi}) yields the
second order approximate solution of (\ref{eq:vartheta2}) in the
following form
\begin{eqnarray}
  \vartheta & = &  \frac{\hat{\beta}}{\mu} + Q \cos(\Zeta +
\tau) +\sqrt{\eps}\frac{\hat{\beta}\,Q}{4}\sin\!\left(Z\right) \\
   &+& \frac{\sqrt{\eps}w\sin\!\left(\tau\right)}{4\sqrt{\mu}}
 - \frac{\sqrt{\eps}\sin\!\left(2\tau
\right)}{3\sqrt{\mu}} + o(\sqrt{\eps})
\end{eqnarray}
which after changes of variable $\theta = - \tau +
\sqrt{\eps}\vartheta$ and parameters $w=\omega^2/\eps$,
$\hat{\beta}=\beta/\sqrt{\eps}$ results in the approximate
solution of the original equation (\ref{eq:theta})
\begin{eqnarray}
  \theta & = & -\tau + \frac{\beta}{\mu} + \sqrt{\eps} Q \cos(\Zeta +
\tau) +\sqrt{\eps}\frac{\beta\,Q}{4}\sin\!\left(Z\right) \nonumber\\
   &+& \frac{\omega^2 \sin\!\left(\tau\right)}{4\sqrt{\mu}}
 - \frac{\eps\sin\!\left(2\tau\right)}{3\sqrt{\mu}} +
 o(\eps).\label{eq:thetaappr2}
\end{eqnarray}
Agreement of solution (\ref{eq:thetaappr2}) with the numerical
experiment is shown in Fig. \ref{f:B001}. We see that the
amplitude of angular velocity oscillations is much higher than
that for not small $\beta$ in Fig. \ref{f:B05}.

\section{Domains of applicability}\label{s:domain} For asymptotic solutions in the previous two sections the parameter
constraints are more strict than those for the existence of stable
exact solution in (\ref{eq:existab}). Thus, for our analysis in
Section \ref{s:notsmall} to be valid we must exclude cases when
$\beta = o(1)$ or/and $\sqrt{\mu^2-\beta^2} = o(1)$. The case when
$\beta = o(1)$ and $\mu=O(1)$ is studied in Section
\ref{s:notsmall}. Note that the case when both $\beta = o(1)$ and
$\mu = o(1)$ meaning $\sqrt{\mu^2-\beta^2} = o(1)$ has already
been studied in the literature, for example in the more general
model of unbalanced rotor in \cite{Blekhman54}. The case when
$\beta = O(1)$ and $\mu = o(1)$ is not feasible for asymptotic
rotational solution. Indeed, with such assumptions the generating
system $\ddot{\theta} + \beta \dot{\theta} = 0$ has only constant
solutions.
\begin{table}
  \centering
  \begin{tabular}{c|c|c|}
     & small $\beta$ &  not small $\beta$\\
    \hline
    small $\mu$ & studied in the literature & no rotations \\
    \hline
    not small $\mu$ & studied in section \ref{s:small} & studied in section \ref{s:notsmall}\\
    \hline
  \end{tabular}\caption{Model assumptions on smallness of dimensionless damping $\beta$ and dimensionless semiaxes half-sum $\mu$
  of the ellipse along which the pivot of the pendulum moves in the problem to find pendulum rotations. In all cases we assume that
  dimensionless half-difference $\eps$ of semiaxes is small as well as $\omega^2$.}\label{t:ma}
\end{table}
These different cases are presented in the Table \ref{t:ma}.

\begin{figure}
  \includegraphics[width=0.5\textwidth]{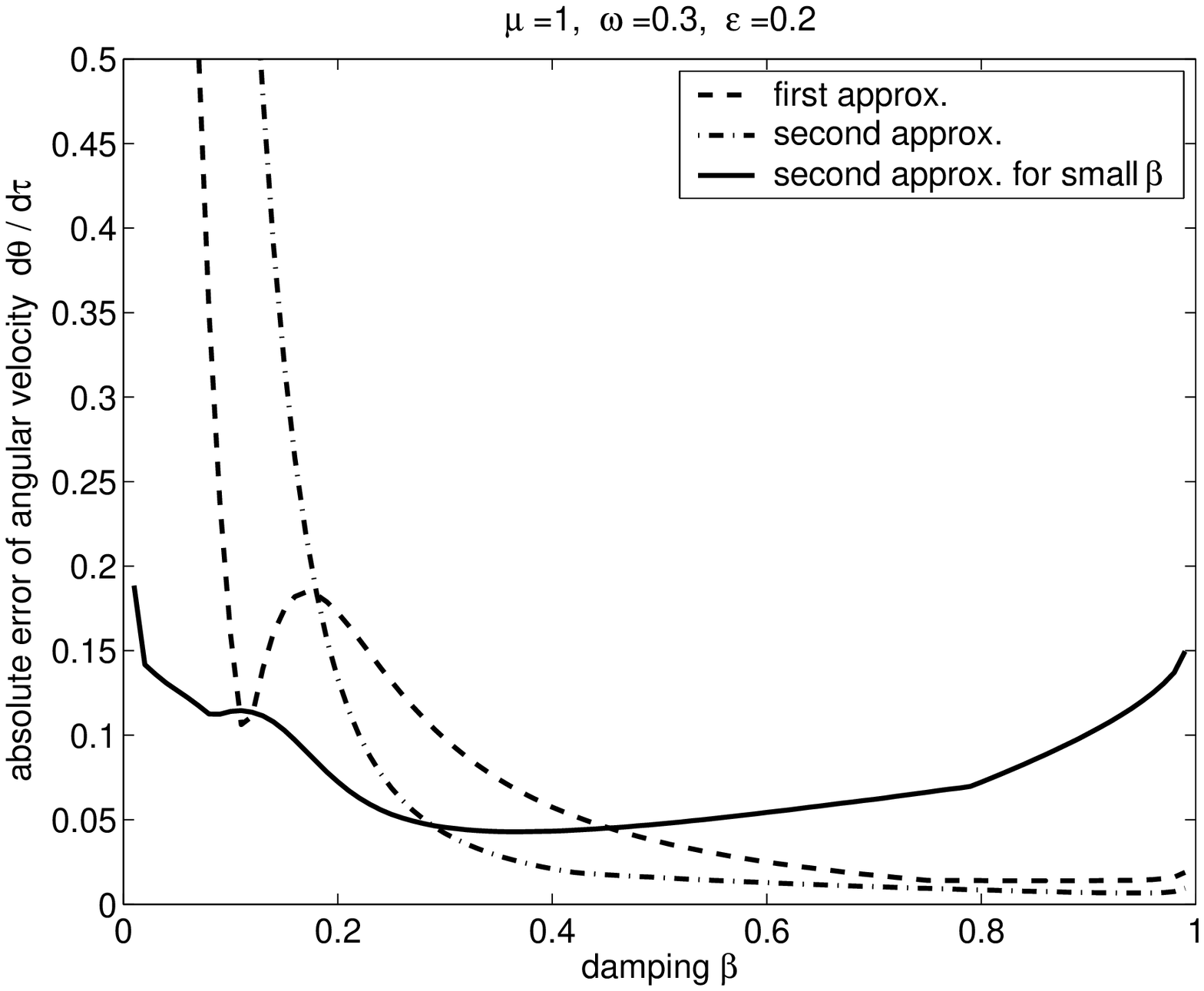}\includegraphics[width=0.5\textwidth]{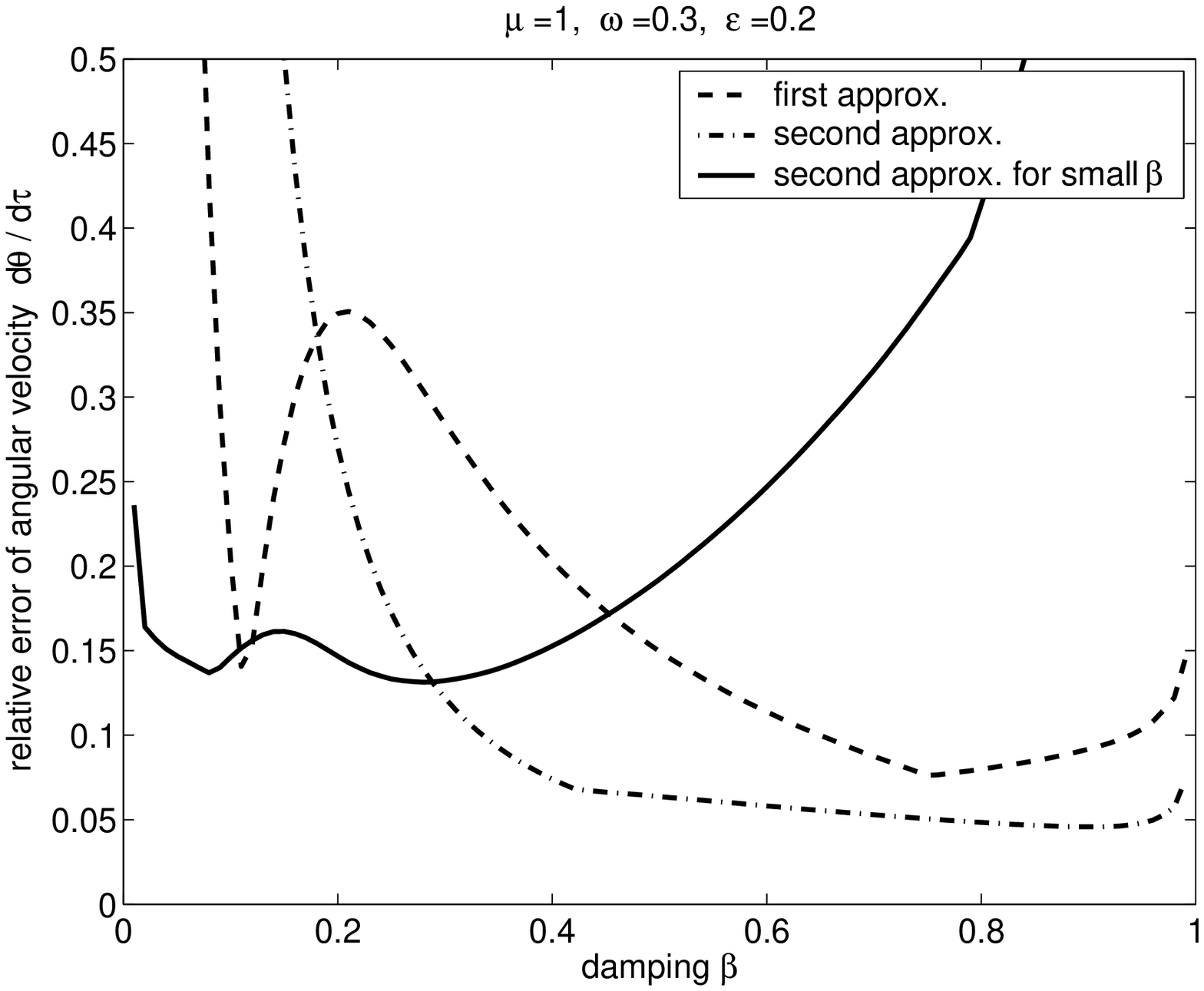}\\
  \caption{Maximal absolute (left) and relative (right) deviation of analytically obtained angular velocities
  $\dot{\theta}$ in (\ref{eq:stheta1}), (\ref{eq:stheta2}), and (\ref{eq:thetaappr2}) from the
  numerically obtained $\dot{\theta}$. Solution (\ref{eq:thetaappr2}) specially obtained for better approximation at small damping
  $\beta$. Error in the right graph is calculated relative to the oscillation
  amplitude of angular velocity $\dot{\theta}$.}\label{f:err}
\end{figure}

To show quantitatively the limits of applicability of the
assumptions in Sections \ref{s:notsmall} and \ref{s:small} we plot
the absolute and relative angular velocity errors depending on
parameter $\beta$ while excitation was constant $\mu = 1$, see
Fig. \ref{f:err}.

\section{Conclusion}
The exact rotational solutions in the case of equal excitation
amplitudes and zero gravity are obtained. The conditions for
existence and stability of such solutions are derived. Based on
these exact solutions the approximate solutions are found both for
high and small linear damping, assuming that the amplitudes of
excitations are not small. Comparison between approximate and
numerical solutions shows a good agrement for the damping values
of the assumed order.

\section*{Acknowledgement} The author would like to thank Professor
Alexander P. Seyranian from Lomonosov Moscow State University for
his valuable comments on the text of the paper.

\section*{Appendix}
Non-uniform second order linear differential equation
\begin{equation}\label{eq:varthetalin}
\ddot{\vartheta} + \beta\dot{\vartheta} +
\sqrt{\mu^2-\beta^2}\vartheta =\A_n\cos\!\left(n\tau\right) +
\B_n\sin(n\tau),
\end{equation}
has solutions which obey the principle of superposition and
contain only harmonics of the right-hand side in
(\ref{eq:varthetalin})
\begin{eqnarray}\label{eq:gensol}
  \vartheta[n] & = & -\frac {(n^2-\sqrt {{\mu}^{2}-\beta^{2}})\A_n + n\beta\B_n
}{(n^2-1)\beta^{2}+{\mu}^{2}+n^2(n^2-2\sqrt{\mu^2-\beta^2})}\cos\!\left(n\tau\right)\nonumber\\
 & - & \frac{-n\beta\A_n + (n^2-\sqrt{\mu^2-\beta^2})\B_n}{(n^2-1)\beta^{2}+{\mu}^{2}+n^2(n^2-2\sqrt{\mu^2-\beta^2})}\sin\!\left(
n\tau \right)
\end{eqnarray}
where $n = 0,1,2,\ldots$. Thus, solution with $n=0$ is the
following
\begin{equation}\label{eq:gensol0}
    \vartheta[0] = \frac{\A_0}{\sqrt{{\mu}^{2}-\beta^{2}}},
\end{equation}
when $n=1$
\begin{eqnarray}\label{eq:gensol1}
  \vartheta[1] & = & -\frac {(1-\sqrt {{\mu}^{2}-\beta^{2}})\A_1 +
  \beta\B_1
}{{\mu}^{2}+1-2\sqrt{\mu^2-\beta^2}}\cos\!\left(\tau\right)\nonumber\\
 & - & \frac{-\beta\A_1 + (1-\sqrt{\mu^2-\beta^2})\B_1}{{\mu}^{2}+1-2\sqrt{\mu^2-\beta^2}}\sin\!\left(
\tau \right)
\end{eqnarray}
when $n=2$
\begin{eqnarray}\label{eq:gensol2}
  \vartheta[2] & = & -\frac {(4-\sqrt {{\mu}^{2}-\beta^{2}})\A_2 +
  2\beta\B_2
}{3\beta^{2}+{\mu}^{2}+4(4-2\sqrt{\mu^2-\beta^2})}\cos\!\left(2\tau\right)\nonumber\\
 & - & \frac{-2\beta\A_2 + (4-\sqrt{\mu^2-\beta^2})\B_2}{3\beta^{2}+{\mu}^{2}+4(4-2\sqrt{\mu^2-\beta^2})}\sin\!\left(
2\tau \right)
\end{eqnarray}

\bibliographystyle{enoc}

\begin{thebibliography}{}

\bibitem{Lenci_Rega} Lenci,~S., Pavlovskaia,~E., Rega,~G., and Wiercigroch,~M.
 Rotating solutions and stability of parametric pendulum by
perturbation method. \emph{Journal of Sound and Vibration}
\textbf{310}, 2008, pp.~243--259.

\bibitem{Xu_Wiercigroch} Xu,~X., Wiercigroch,~M. Approximate analytical solutions for oscillatory and rotational
motion of a parametric pendulum. \emph{Nonlinear Dyn}.
\textbf{47}, 2007, pp.~311--320.

\bibitem{Bog_Mitr} Bogolyubov,~N.~N., Mitropol'skii,~Yu.~A. {\em Asymptotic
Methods in the Theory of Nonlinear Oscillations}. Gordon and
Breach, New York, 1961.

\bibitem{Seyran} Seyranian,~A.~P., Yabuno,~H., Tsumoto,~K. Instability and periodic motion of a physical pendulum
with a vibrating suspension point (theoretical and experimental
approach). {\em Doklady Physics}, \textbf{50}(9), 2005,
pp.~467--472.

\bibitem{Horton}
Horton~B., Sieber~J., Thompson,~J.~M.~T., Wiercigroch,~M. Dynamics
of the elliptically excited pendulum. {\em arXiv:0803.1662v1
[math.DS]} 11 Mar 2008.

\bibitem{Blekhman54} Blekhman,~I.~I. Rotation of an unbalanced rotor caused by harmonic oscillations of
its axis. \emph{Izv. AN SSSR, OTN}, \textbf{8}, 1954, pp.~79--94.
(in Russian)

\bibitem{Blekhman79} Blekhman,~I.~I.: \emph{Vibrations in Engineering}. A Handbook. Vol. 2. Vibrations of
Nonlinear Mechanical Systems, Mashinostroenie, Moscow, 1979. (in
Russian)

\bibitem{Blekhman} Blekhman,~I.~I. {\em Vibrational Mechanics. Nonlinear Dynamic Effects, General Approach, Applications,}
World Scientific, Singapore 2000, 509 pp.

\bibitem{Akulenko} Akulenko,~L.~D. Higher-order averaging schemes in the theory of non-linear
oscillations. \emph{J. Appl. Mafhs Mechs}, \textbf{65}(5), 2001,
pp.~817--826.

\bibitem{Trueba} Trueba,~J.~L., Baltan\'{a}as,~J.~P., Sanju\'{a}an,~M.~A.~F. A generalized perturbed
pendulum. \emph{Chaos, Solitons and Fractals}. \textbf{15}, 2003,
pp.~911–-924.

\bibitem{Fidlin} Fidlin,~A., Thomsen,~J.~J.
Non-trivial effects of high-frequency excitation for strongly
damped mechanical systems. {\em International Journal of
Non-Linear Mechanics}, \textbf{43}, Issue 7, 2008, pp.~569--578.

\bibitem{Belyak} Belyakov,~A.~O., Seyranian,~A.~P. The Hula-Hoop
Problem. {\em Doklady Physics}, \textbf{55}(2), 2010, pp. 99--104.

\bibitem{Nayfeh} Nayfeh,~A.~H. Perturbation Methods. \emph{Wiley-Interscience, New York},
1973, 425~pp.

\bibitem{Vol_Morg} Volosov,~V.~M., Morgunov,~B.~I. {\em
\emph{Averaging Method in the Theory of Nonlinear Oscillatoratory}
Systems}. MSU, Moscow, 1971.


\end{thebibliography}


\end{document}